\documentclass[10pt]{iopart}
\usepackage[super,compress]{cite}
\usepackage{graphicx}
\usepackage{subfigure}
\expandafter\let\csname equation*\endcsname\relax
\expandafter\let\csname endequation*\endcsname\relax
\usepackage{amsmath}
\usepackage{mathtools}
\usepackage{iopams}
\usepackage{amssymb}
\usepackage{breakurl}
\usepackage{url,hyperref}
\begin{document}

\title[Apiwit Kittiratpattana]{Modeling (anti)deuteron formation at RHIC with a geometric coalescence model
}
\author{A Kittiratpattana$^1$, M F Wondrak$^{2,3}$, M Hamzic$^4$, M Bleicher$^{2,3,5}$, A Limphirat$^1$, C Herold$^1$}
\address{$^1$ Center of Excellence in High Energy Physics \& Astrophysics, \\ Suranaree University of Technology, Nakhon Ratchasima 30000, Thailand}
\address{$^2$ Helmholtz Research Academy Hesse for FAIR (HFHF), Campus Frankfurt, \\
Max-von-Laue-Str. 12, 60438 Frankfurt am Main, Germany}
\address{$^3$ Institut f\"ur Theoretische Physik, Johann Wolfgang Goethe-Universit\"at, \\
Max-von-Laue-Str. 1, 60438 Frankfurt am Main, Germany}
\address{$^4$ University of Sarajevo, Obala Kulina bana 7/II, \\ 71000 Sarajevo, Bosnia and Herzegovina}
\address{$^5$ GSI Helmholtzzentrum f\"ur Schwerionenforschung GmbH,\\
Planckstr. 1, 64291 Darmstadt, Germany}

\ead{apiwitkitti@g.sut.ac.th}
\vspace{10pt}

\begin{abstract}
We study (anti)deuteron formation rates in heavy-ion collisions in the framework of a wave-function based coalescence model. The main feature of our model is that nucleons are emitted from the whole spherically symmetric fireball volume, while antinucleons are emitted only from a spherical shell close to the surface. In this way, the model accounts for nucleon-antinucleon annihilations in the center of the reaction at lower beam energies. Comparison with experimental data on the coalescence parameter in the range $\sqrt{s_{NN}}= 4.7-200$~GeV allows us to extract radii of the respective source geometries. Our results are qualitatively supported by data from the UrQMD transport model which shows a comparable trend in the geometric radii as a function of beam energy. In line with our expectations, we find that at lower energies, the central region of the fireball experiences stronger annihilation than at higher energies.
\end{abstract}

\vspace{2pc}

\submitto{\PS}
\maketitle
\ioptwocol

\section{Introduction}

Relativistic Heavy-ion Collision experiments such as ALICE, PHENIX, and STAR, enable us to probe nuclear matter under extreme conditions by colliding heavy nuclei at high energies. Such extreme conditions, i.e., high temperatures and/or high baryon densities \cite{sarkar2009physics}, can mimic the conditions present in the early universe or in neutron star mergers \cite{Rufa:1990zm}. The production of light (anti)nuclei in these systems has been under investigation from astrophysics to nuclear particle physics in the last years \cite{DeGrand:1984by,Zhu:2015voa,Hagedorn:1960zz,Butler:1961pr,Nagle:1996vp,SchaffnerBielich:2000wj,Monreal:1999mv,Chen:2003ava,Oh:2007vf,Chen:2018tnh}. The abundance of light (anti)nuclei could provide us with information about why our universe consists of matter rather than antimatter \cite{Malaney:1993ah}. In this work, we focus on the formation of (anti)deuterons which is the simplest nucleon-nucleon composite particle.


To understand and predict the formation rate of light nuclei, a most rigorous description would be based upon quantum chromodynamics (QCD), which explains the dynamics of colour exchange between individual quarks at a microscopic level. Unfortunately, it is extremely difficult today to describe such a dynamic system at strong coupling by first principles \cite{Beane:2011iw,Beane:2012vq,Yamazaki:2012hi}. However, one can perform effective microscopic calculations based on the sudden approximation, perturbation theory, and transport models to calculate the (anti)deuteron formation rate \cite{Kapusta:1980zz, Aaij:2012hu, Bondorf:1978kz}.  The two most popular models are the thermal model \cite{DasGupta:1981xx,Mekjian:1978zz,Mrowczynski:2009wk} which treats each (composite) particle species as a unique degree of freedom in a statistical Fermi/Bose gas approach, and the coalescence model, i.e., formation from localized (anti)nucleons in phase-space coordinates   \cite{Mrowczynski:2016xqm, Steinheimer:2012tb, Jennings:1982wi, Aichelin:1987rh}. They provide fundamentally different interpretations of the deuteron formation process \cite{Mrowczynski:2020ugu}.


The simple momentum space coalescence model states that once particles are carrying similar momenta, they will coalesce and form into a new composite particle, e.g., a deuteron from two nucleons. This explanation has been successful for decades. However, deviations from this simple approach have been reported in 1992, with the failure in the prediction of the antideuteron production in Si-Au collisions at AGS \cite{Aoki:1992mb} and have later led to the development of the phase-space coalescence approach. The reason for the failure to describe anti-deuterons was pinned down to the fact that the deuteron and antideuteron formation rates are sensitive to the source shape \cite{Sato:1981ez, Leupold:1993ms}. In 1993, Mrówczyński included the spatial geometry of the collision fireball into the coalescence model \cite{Mrowczynski:1993cx}. The main assumption in his approach is that antinucleons are emitted from the outer shell of the source due to a high probability of nucleon-antinucleon annihilation in the central baryon-rich region, while nucleons are emitted from the entire volume. The model assumes a spherical symmetry and successfully describes experimental data from Si+Au collisions at $E_{lab}=14.6A$~GeV.


Nowadays, with more experimental data available on deuteron and antideuteron formation rates at different energies, we aim to improve Mr\'owczy\'nski's coalescence model to extract the spatial geometries of the nucleon and antinucleon source at each beam energies from $\sqrt{s_{NN}}=7.7-200$ GeV. It would be helpful to know how the source geometries behave as a function of energy so that we could explain why and how (anti)deuterons, and ultimately light (anti)nuclei, are produced. Firstly, we apply Mr\'owczy\'nski's coalescence model directly to the energy dependence of (anti)nucleon source geometry calculations. Here, any suppression in the nucleon source is neglected. Then in a second method, we constrain the root-mean-squared radius of the nucleon source via the number of charged particles. With this second approach, both nucleon and antinucleon sources face suppression at the central core.


Recently, deuteron production from the coalescence model has been studied within the Ultra-Relativistic Quantum Molecular Dynamics model (UrQMD) \cite{ Bleicher:2019oog, Hillmann:2019wlt} (see also \cite{Nagle:1994wj} for previous RQMD studies). The UrQMD transport model \cite{Bass:1998ca,Bleicher:1999xi} is a microscopic transport model based on the binary scattering of hadrons, resonance excitations, resonance decays, and string dynamics, as well as strangeness exchange reactions \cite{Graef:2014mra}. The deuteron rapidity and transverse momentum distributions as well as the $d/p$ and $\bar{d}/\bar{p}$ ratios are in agreement with experimental data \cite{Sombun:2018yqh}. In contrast, the model presented in this paper, based on Mr\'owczy\'nski's coalescence model, assumes instantaneous emission. To further support our idea, we apply UrQMD to study how nucleons and antinucleons freeze-out geometrically. The results are then compared with the coalescence model.


\section{(Anti)Nucleon Source to (Anti)Deuteron Formation Rate}

Our study applies the Mr\'owczy\'nski model to extract information about the spatial distribution of (anti)deuterons from the experimental data. The (anti)deuteron distributions and (anti)nucleon distributions now depend on the adjustable emission-source parameters which can be related to the initial conditions. In this section, we will follow the description from Kittiratpattana et al. (2020) \cite{Kittiratpattana:2020daw} and Mr\'owczy\'nski (1993) \cite{Mrowczynski:1993cx}.

The main assumption is that antinucleons produced near the center of the interaction region of the collision have a very high probability to experience secondary interactions and annihilate in the baryonic environment. On the other hand, the antinucleons produced on the outer surface can survive and leave the fireball to coalesce. Nucleons are freely emitted from the whole source, i.e., fireball volume. From these different shapes of anti-nucleon and nucleon sources, we aim at explaining the differences between their formation rates. The invariant cross section of (anti)deuteron formation can be written as,
    \begin{equation}
        E\left(\frac{d\sigma_d}{d^3\textbf{P}}\right) = B_2\left(E\frac{d\sigma_{p}}{d^3\textbf{P}/2}\right)^2~,
    \end{equation}
where $B_2$ is the coalescence parameter which can be measured in experiment. We hereby neglect n-p correlations, as well as their mass difference, thus, $\textbf{p}_p = \textbf{p}_n = \textbf{P}/2$. The coalescence parameter $B_2$ is linked to the spatial coalescence formation rate via $\mathcal{A}\equiv \frac{m_N}{2}B_2$, where $m_N=0.94$~MeV is the nucleon mass. In the particle source rest frame, the deuteron formation rate is $\mathcal{A}'=\gamma\mathcal{A}$ with $\gamma$ being the Lorentz factor of the deuteron motion. The formation rate is
    \begin{equation}
        \mathcal{A} = \frac{3}{4}(2\pi)^3\int d^3r_1d^3r_2\mathcal{D}(\textbf{r}_1)\mathcal{D}(\textbf{r}_2)|\psi_d(\textbf{r}_1,\textbf{r}_2)|^2~.
        \label{eq:A}
    \end{equation}
The source function $\mathcal{D}(\textbf{r})$ describes the probability of finding a nucleon at a given point $\textbf{r}$ at kinetic freeze out. $\psi_d(\textbf{r}_1, \textbf{r}_2)$ denotes the deuteron wave function. The absence of the time variable in this formula, however, does not mean that we neglect dynamical properties, but study a simultaneous emission \cite{Mrowczynski:1993cx}.

Following this ansatz, the nucleon emission source is distributed over the whole fireball, while the antinucleon source around the center of the fireball is suppressed, leading to a surface type emission for antinucleons. The nucleon source function $\mathcal{D}(\textbf{r})$ is parametrized by a normalized Gaussian,
    \begin{equation}
        \label{eq:D}
        \mathcal{D}(\textbf{r}_i) = \frac{1}{(2\pi)^{3/2} r_0^3}\exp{\left( - \frac{\textbf{r}_i^2}{2r_0^2} \right)}~,
    \end{equation}
with $r_0$ representing the source radius which is equivalent to the fireball radius. The antinucleon source function is also normalized and reads,
    \begin{equation}
        \label{eq:Dbar}
            \bar{\mathcal{D}}(\textbf{r}_i) = \left\{
                \begin{aligned}
                    & \begin{aligned}[b]
                        \frac{1}{{(2\pi)^{3/2} (r_0^3-r_*^3)}}\left[ \exp{\left( -\frac{\textbf{r}_i^2}{2r_0^2} \right)}\right. \\
                         \hspace{-1cm}-\left.\exp{\left( -\frac{\textbf{r}_i^2}{2r_*^2} \right)}\right],
                      \end{aligned} && \text{for $r_* \neq r_0$.} \\[2ex]
                    & \frac{1}{3(2\pi)^{3/2}}\frac{\textbf{r}_i^2}{r_*^5}     \exp{\left( -\frac{\textbf{r}_i^2}{2r_*^2} \right)}, &&\text{for $r_* = r_0$.}
                \end{aligned}
                    \right.
    \end{equation}
Here, a second Gaussian of width $r_*$ cuts out the central region defining the nucleon-antinucleon annihilation volume where the emission of antinucleons is strongly depleted.
            
The integral in ~\eqref{eq:A} can be formulated using center-of-mass coordinates $\textbf{P}= \textbf{p}_1+\textbf{p}_2$, $\textbf{R}=\frac{1}{2}\left( \textbf{r}_1+\textbf{r}_2\right)$ and relative coordinates  $\textbf{q}=\frac{1}{2}(\textbf{p}_1-\textbf{p}_2$), $\textbf{r}=\textbf{r}_1-\textbf{r}_2$. The deuteron wave function then reads 
            \begin{equation}
              \psi_d(\textbf{r}_1, \textbf{r}_2) = \text{e}^{i\textbf{P}\cdot \textbf{R}}\phi_d(\textbf{r})~,
            \end{equation}
with the Hulth\'en wave function,
            \begin{equation}
            \label{eq:wavefn}
              \phi_d(\textbf{r}) = \left( \frac{\alpha\beta(\alpha+\beta)}{2\pi(\alpha-\beta)^{2}}\right)^{1/2}\frac{\text{e}^{-\alpha r}-\text{e}^{-\beta r}}{r}~,
            \end{equation}
where $\alpha=0.23$~fm$^{-1}$ and $\beta = 1.61$~fm$^{-1}$ \cite{hodgson1971nuclear}. The formation rate $\mathcal{A}$ in relative coordinates reads
            \begin{equation}
            \label{eq:Ar}
              \mathcal{A} \equiv \frac{3}{4}(2\pi)^3 
              \int d^3 r \mathcal{D}_r(\textbf{r}) |\phi_d(\textbf{r})|^2~,
            \end{equation}
 the nucleon source function becomes
            \begin{equation}
            \label{eq:Dr}
              \mathcal{D}_r( \textbf{r} ) =  \frac{1}{(4\pi)^{3/2}r_0^3}\exp{ \left( -\frac{\textbf{r}^2}{4r_0^2} \right)}~,
            \end{equation}
and the antinucleon source function for $r_0=r_*$ reads,
            \begin{equation}
            \begin{aligned}
            \label{eq:Drbar1}
                \bar{\mathcal{D}}_r(\textbf{r}) = &4\pi\left( \frac{1}{3 (2\pi)^{3/2} r_{*}^5}\right)^2 \left[ \frac{15\sqrt{\pi}}{16}r_{*}^{7} \right.\\
                &\left.+\frac{6\sqrt{\pi}\textbf{r}^2}{16}r_{*}^{5} +\frac{\sqrt{\pi}\textbf{r}^4}{64}r_{*}^{3}\right]\exp\left(-\frac{\textbf{r}^2}{2r_{*}^2}\right)~.
            \end{aligned}
            \end{equation}
and for $r_0 \neq r_*$,
            \begin{equation}
            \begin{aligned}
            \label{eq:Drbar2}
                \bar{\mathcal{D}}_r(\textbf{r}) = &\frac{1}{(4\pi)^{\frac{3}{2}} (r_{0}^3-r_{*}^3)^2} \left[ r_{0}^3 \exp{\left(-\frac{\textbf{r}^2}{4r_{0}^2}\right)}  + r_{*}^3 \exp{\left(-\frac{\textbf{r}^2}{4r_{*}^2}\right)} \right. \\ 
                &\left. - \tfrac{2^{\frac{5}{2}}r_{0}^3 r_{*}^3}{(r_{0}^2+r_{*}^2)^{\frac{3}{2}}}\exp{\left(-\frac{\textbf{r}^2}{2\left(r_{0}^2+r_{*}^2\right)}\right)}\right]~.
            \end{aligned}
            \end{equation}
            
\begin{figure}
    \centering
    \includegraphics[width=0.9\columnwidth]{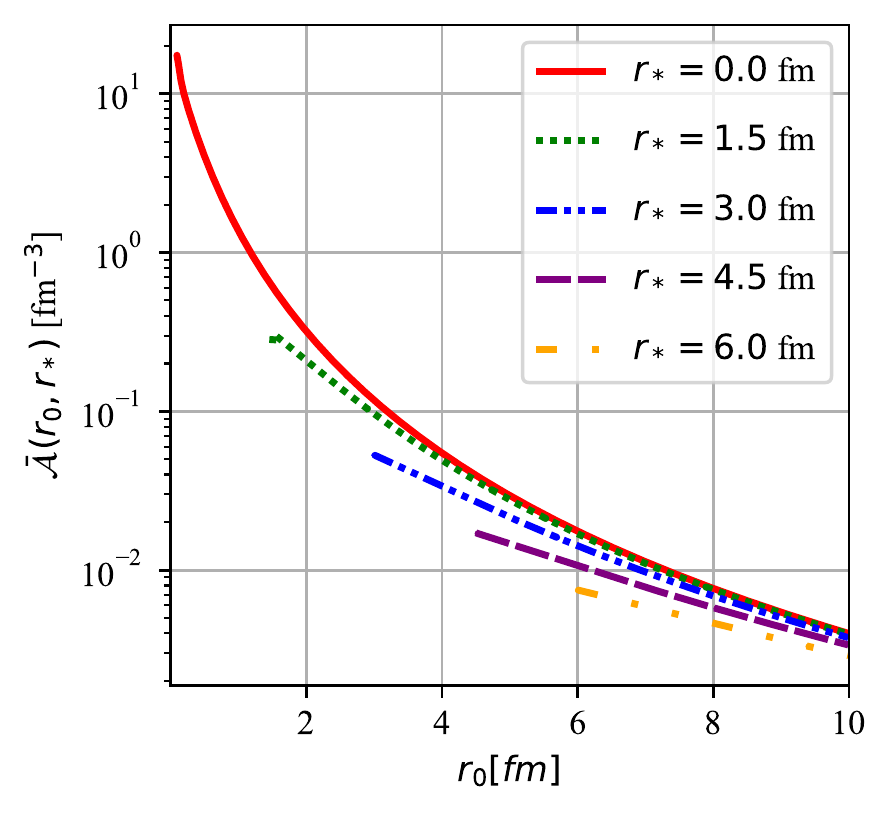}
    \caption{The antideuteron formation rate as a function of $r_0$ with different values of $r_*$ starting from $r_*=0-6$~fm. The smaller suppression radius $r_*$ gives the higher formation rate $\bar{\mathcal{A}}(r_0, r_*)$ as the deuteron or antideuteron formation rate with $r_*$ sets an upper limit to the plot \cite{Kittiratpattana:2020daw}.}
    \label{fig:fig1}
\end{figure}

Figure \ref{fig:fig1} shows the antideuteron formation rate $\bar{\mathcal{A}}(r_0, r_*)$ according to ~\eqref{eq:Ar} with various suppression radii $r_*$. The deuteron formation rate is given by the red solid line with $r_*=0$~fm. Note that the region where the antideuteron formation $\bar{\mathcal{A}}(r_0,r_*)$ vanishes is where the suppression radius becomes larger than the fireball radius, i.e., $r_*>r_0$, which contradicts our physical interpretation of $r_0$ and $r_*$.

\begin{figure}
    \centering
    \includegraphics[width=0.9\columnwidth]{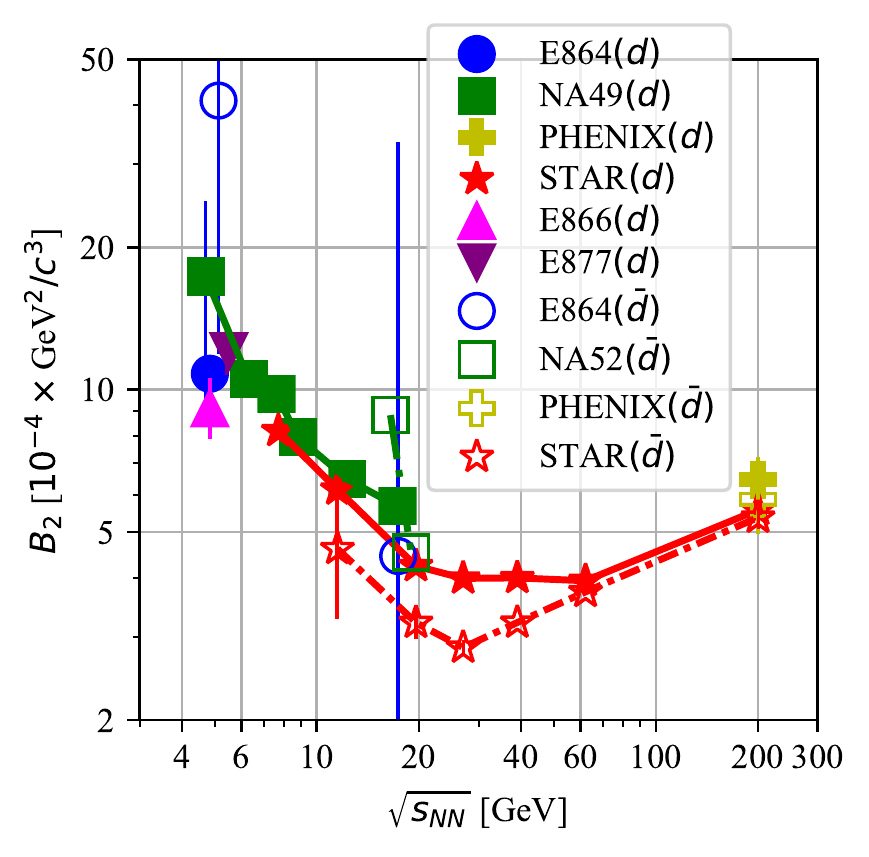}
    \caption{The energy dependence of coalescence parameter $B_2$ of deuteron $(d)$ and antideuteron $(\bar{d})$ from several experiments ranging from $\sqrt{s_{NN}}=4.7-200$~GeV \cite{Ambrosini:1998wg,VanBuren:1999ir,Armstrong:2000gd,Adam:2019wnb}.}
    \label{fig:fig2}
\end{figure}

To extract the (anti)nucleon source geometry, we fit to experimental coalescence data $B_2$ from low- to mid-energy range experiments as shown in figure \ref{fig:fig2}. We use data from NA49$(d)$ and STAR$(d)$ for extracting the nucleon source geometry, and from STAR$(\bar{d})$ for evaluating the antinucleon source geometry $(r_0, r_*)$. First, the fireball radius $r_0$ is extracted from $B_2$ of NA49 and STAR$(d)$ data since the antinucleon source also shares the same fireball radius. Then, the suppression radius $r_*$ is evaluated from $B_2$ of STAR$(\bar{d})$.

\begin{figure}
    \centering
    \includegraphics[width=0.9\columnwidth]{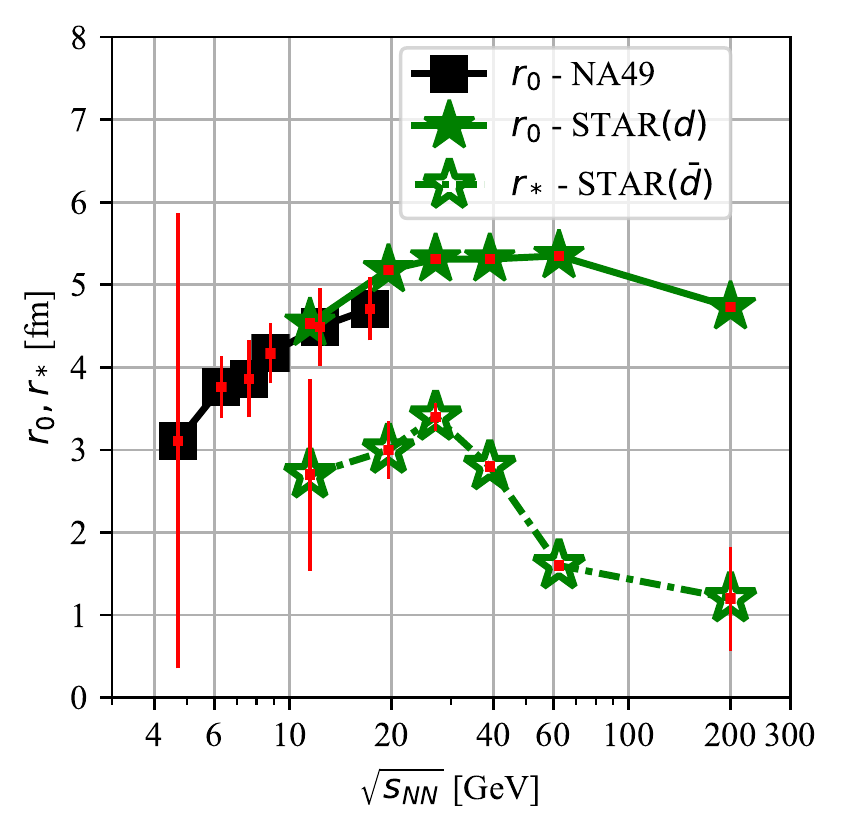}
    \caption{The (anti)nucleon source geometry as a function of energy. The black square solid-line and green star solid-line represent the nucleon fireball radius $r_0$ from NA49$(d)$ and STAR$(d)$ respectively. The green star dashed-line shows the antinucleon suppression radius $r_*$ from STAR$(\bar{d})$. \cite{Kittiratpattana:2020daw}}
    \label{fig:fig3}
\end{figure}

The energy-dependent (anti)nucleon source radii are shown in figure \ref{fig:fig3}. Both fireball size and suppression region increase with energy until $\sqrt{s_{NN}}\simeq 27$~GeV and decrease for higher energies. This could be understood by the change of the dominant particle species in the central region of the fireball at high energies. Since fewer primary scatterings between antinucleons and nucleons occur,  the nucleon-antinucleon annihilation probability would become smaller reflecting a smaller suppression radius in the antinucleon source. Another interesting behaviour is that both fireball and suppression radii seem to reach a maximum at $\sqrt{s_{NN}}=27$~GeV. This could be a sign of a change in the QCD equation of state or of a nutcracker shell-like structure as pointed out by Shuryak   \cite{Shuryak:1998ab}. Such considerations, however, are beyond the scope of this paper.

\section{Charged Volume Constraint}
The above results give rise to various questions. Is the nucleon source at higher energies suppressed by pion domination although the parametrization of the source function of nucleons does not consider any suppression like the antinucleon source? The suppression feature in the nucleon source may be hidden and the extracted fireball radius $r_0$ might be a root-mean-squared radius, RMS-radius, or $\langle r^2 \rangle^{1/2}$, instead. Hence this might underestimate the real geometries. Furthermore, the fireball size should grow with the energy \cite{Aamodt:2011mr}. One usually assumes that the number of charged particles $N_{ch}$ is related to the fireball volume $V_{ch}$ as dictated by thermodynamics. Thus, the fireball volume should always increase with the energy as does $N_{ch}$. In this paper, however, the fireball radius $r_0$ is equivalent to the nucleon source radius. So the decreasing of a fireball $r_0$ at high energies from figure \ref{fig:fig3} needs more investigation.

\begin{figure}
    \centering
    \includegraphics[width=0.9\columnwidth]{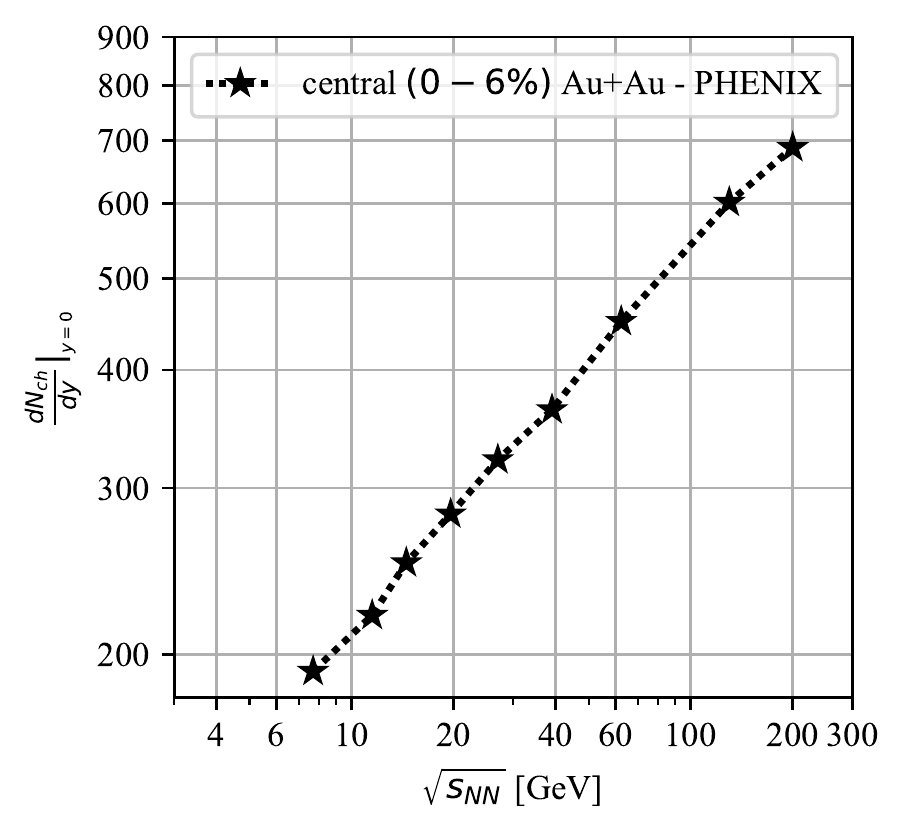}
    \caption{The energy dependence of mid-rapidity charged particle from PHENIX ranging from $\sqrt{s_{NN}}=7.7, 11.5, 14.5, 19.6, 27, 39, 62.4, \text{and}, 200$~GeV \cite{Adare:2015bua}.}
    \label{fig:fig4}
\end{figure}

For these reasons, we constrained our RMS-radius using the number of charged particles at mid-rapidity measured by PHENIX \cite{Adare:2015bua}, see figure \ref{fig:fig4}. 

\begin{figure}%
    \centering
    \includegraphics[width=1.03\columnwidth]{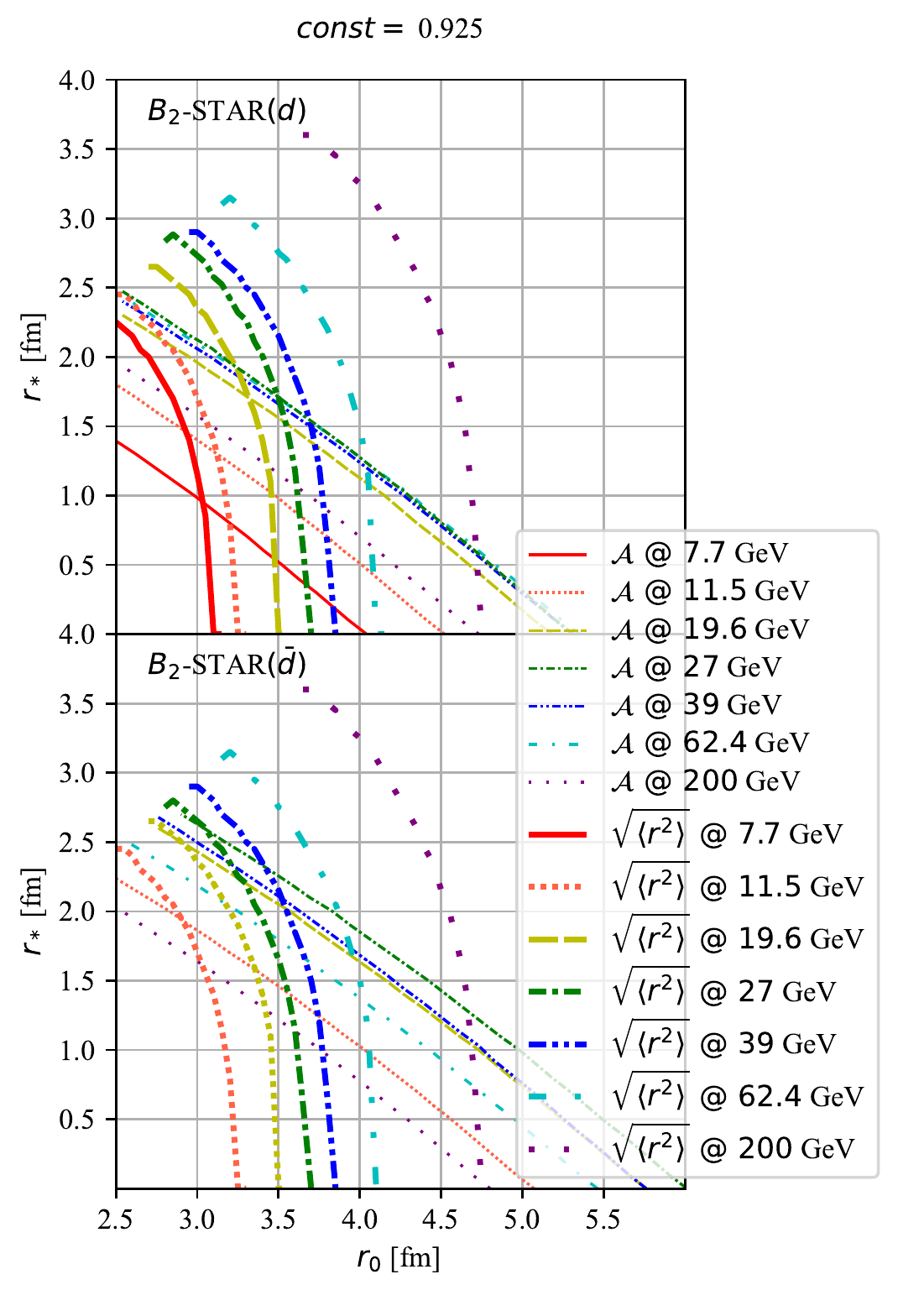}%
    \caption{The (anti)deuteron formation rate $\mathcal{A}(r_0, r_*)$ from STAR$(d)$ and STAR$(\bar{d})$ and the RMS-radius of (anti)nucleon source in the $(r_0, r_*)$ space from PHENIX. The solid-line represents one satisfied the formation rate function. The dotted-line is the $(r_0, r_*)$ satisfied the RMS-radius function.}%
    \label{fig:fig5}%
\end{figure}

In this method, the RMS-radius growth will be consistent with the charged fireball volume $\langle r^2 \rangle^{1/2} = \text{const}\cdot N_{ch}^{1/3}$. Furthermore, the fireball radius $r_0$ and the suppression radius  $r_*$ now will independently grow according to the RMS-radius. The RMS-radius for the (anti)nucleon source reads as,
\begin{equation}
\begin{aligned}
    \langle r^2(r_0, r_*) \rangle^{1/2} &= \left(\int \textbf{r}^2 \bar{\mathcal{D}}(\textbf{r}) d^3r\right)^{1/2} \\
    &= \begin{cases} \sqrt{3}\left(\frac{r_0^5-r_*^5}{r_0^3-r_*^3}\right)^{1/2} , &\text{for $r_0\neq r_*$}\\
    \sqrt{\frac{5}{3}}r_*, &\text{for $r_0 = r_*$}~.
    \end{cases}
    \label{eq:rms-N}
\end{aligned}
\end{equation}
The formation rate $\bar{\mathcal{A}}(r_0, r_*)$ is fitted with the measured coalescence parameter $B_2$ from NA49 and STAR experiment via $\mathcal{A}\equiv \frac{m_N}{2}B_2$. Now the radius of the (anti)nucleon source is the $(r_0, r_*)$ pair that satisfies both $\langle r^2 \rangle^{1/2} = \text{const}\cdot N_{ch}^{1/3}$ and $\mathcal{A}\equiv \frac{m_N}{2}B_2$ as shown in figure \ref{fig:fig5}. According to the $B_2$ data provided by STAR$(d)$, we can study the nucleon source geometry from $\sqrt{s_{NN}}= 7.7-200$~GeV. For antinucleons, the energy provided by STAR$(\bar{d})$ ranges from $\sqrt{s_{NN}}= 11.5-200$~GeV \cite{Adam:2019wnb}. Here, the constant value for the proportionality between (anti)nucleon RMS-radius $\langle r^2 \rangle^{1/2}$ and cube root number of charges particle $N_{ch}^{1/3}$ is found to be $0.925$.

\begin{figure}
    \centering
        \includegraphics[width=0.9\columnwidth]{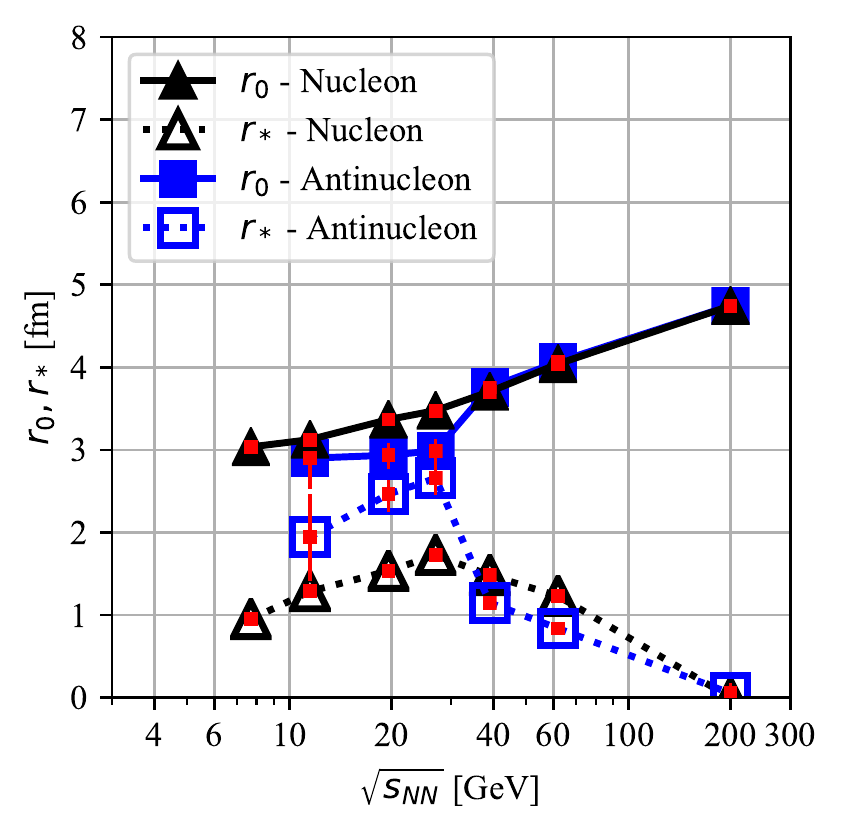}
    \caption{The (anti)nucleon source geometry plot with a function of energy $\sqrt{s_{NN}}$. The solid-line and dashed-line indicate the fireball radius $r_0$ and suppression radius $r_*$ respectively. The black triangle symbol and blue square symbol differentiate the source type between nucleon and antinucleon source.}
    \label{fig:fig6}
\end{figure}

The crossing points between formation rate $\mathcal{A}$ and the RMS-radius $\langle r^2\rangle^{1/2}$ determine the (anti)nucleon source geometries and are shown in figure \ref{fig:fig6}. We found that both nucleon and antinucleon sources grow with the energy as we expect given the constraint. Especially for $\sqrt{s_{NN}} \geq 39$~GeV, both nucleon and antinucleon sources seem to be sharing the same fireball radii as we speculated in the previous sections. The increase of the suppression region $r_*$ agrees with the previous result (figure \ref{fig:fig3}) where it reaches a maximum at $\sqrt{s_{NN}}=27$~GeV. Furthermore, as the energy increases, both source radii rapidly decrease. The reasoning for this is similar to our former statement of the pion enhancement. At higher energy, pion dominates the fireball and suppress nucleon-antinucleon annihilation. However, the drop of $r_*$ to zero at $\sqrt{s_{NN}}=200$~GeV means that the pion domination does not directly suppress the nucleon source. 

\section{Freeze-out Geometries from UrQMD}
The two previous geometries are in agreement with the depleting suppression of antinucleon sources at higher energies.  However, we want to see how exactly the pion enhancement affects both the nucleon and antinucleon distribution at freeze-out at different energies regardless of the coalescence parameter $B_2$.

\begin{figure}%
    \begin{subfigure}[$\sqrt{s_{NN}}=11.5$~GeV]{
    \includegraphics[width=0.9\columnwidth]{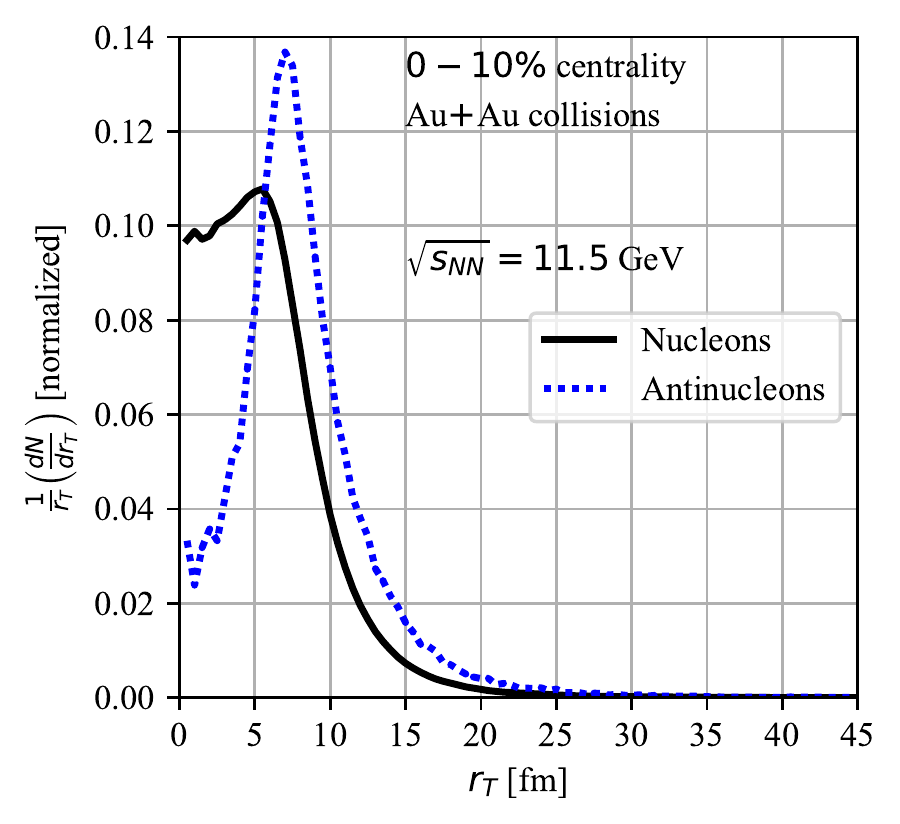}}
    \end{subfigure}
    \begin{subfigure}[$\sqrt{s_{NN}}=11.5$~GeV]{\includegraphics[width=0.9\columnwidth]{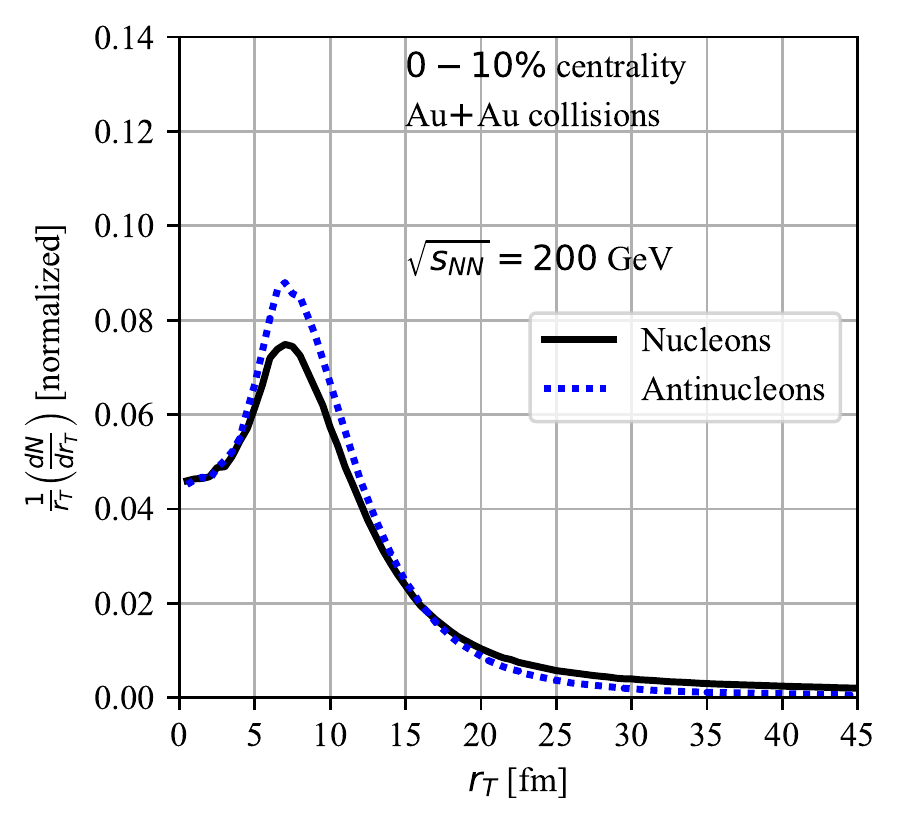}}
    \end{subfigure}
    \caption{The transverse distribution of (anti)nucleons as a function of energy at mid-rapidity Au+Au $0-10\%$ collisions. The solid-line shows the nucleon distributions and dashed-line is for the antinucleons. The plots are normalized to the unity \cite{Kittiratpattana:2020daw}.\\
    }%
    \label{fig:fig7}%
\end{figure}

We use UrQMD to simulate Au+Au  collisions at $0-10\%$ centrality at $\sqrt{s_{NN}}=7.7$, $11.5$, $14.5$, $19.6$, $27$, $39$, $62.4$, and $200$~GeV. The freeze-out coordinates of nucleons and antinucleons can then be extracted directly. The freeze-out geometries will be examined in the transverse plane and the transverse radius $r_T$ distributions of (anti)nucleons will be used, $\frac{1}{r_T}\left(\frac{dN}{dr_T} \right)$\footnote{We take the approximated cylindrical geometry into account. Thus we divide by $r_T$ to regulate the distributions in the center-of-mass frame}. We show examples for the (anti)nucleon distributions as a function of energy at $\sqrt{s_{NN}}=11.5$ and  $200$~GeV in figure \ref{fig:fig7}. They represent the low and high energy behaviour of the (anti)nucleon distributions. 

We directly extract the (anti)nucleon source geometry by fitting the source functions $\mathcal{D}(\mathbf{r})$ and $\bar{\mathcal{D}}(\mathbf{r})$ to these distributions. Note that for large $r_T > 15$~fm, the antinucleon's tail lies below the nucleon's one which ensures proper normalization. At $\sqrt{s_{NN}}=11.5$~GeV, figure~\ref{fig:fig7}(a), antinucleons are clearly suppressed at the core of the fireball ($r_T < 5$~fm). Those which survive freeze out on a shell-like structure. The nucleon distribution seems to be constant before smearing out towards the surface. On the other hand, at $\sqrt{s_{NN}}=200$~GeV, figure~\ref{fig:fig7}(b), we can clearly see that both antinucleons and nucleons are showing a suppression effect at the central region (almost the same at $r_T \simeq 0$~fm). This is in contradiction to the result from the charged volume constraint method where the suppression region vanishes, i.e., $r_* \simeq 0$~fm, at $200$ GeV for both species.  

\begin{figure}
    \centering
    \includegraphics[width=0.9\columnwidth]{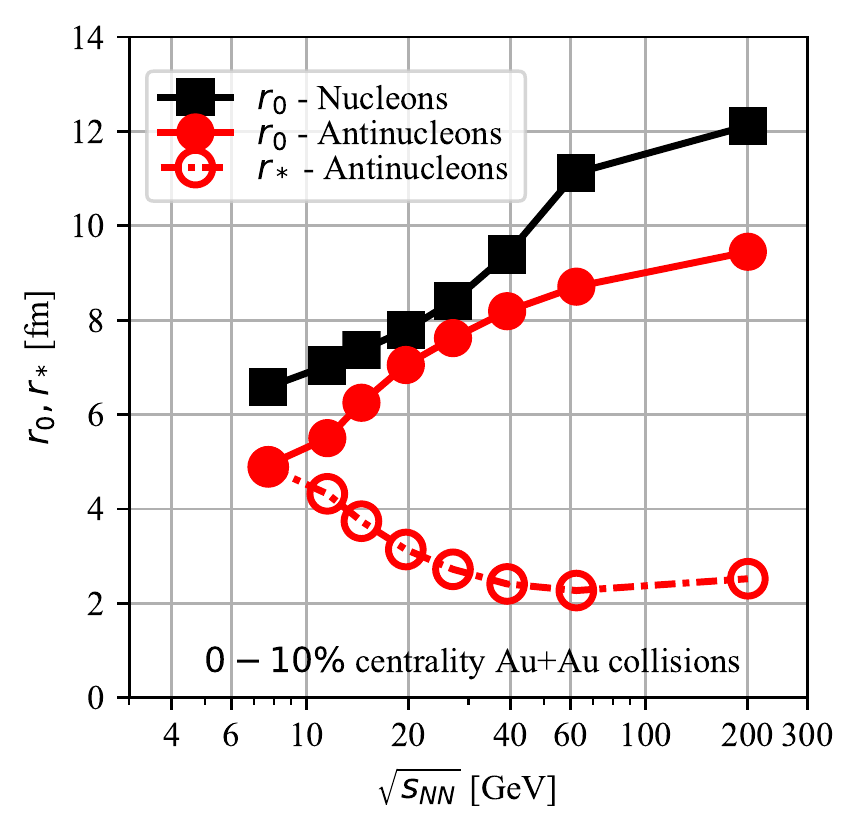}
    \caption{The (anti)nucleon source geometry at central ($0-10\%$) Au+Au collisions (UrQMD) as a function of center-of-mass energy from $\sqrt{s_{NN}}=7.7-200$~GeV. The black square symbol represents the nucleon source. The red cycle symbol represents the antinucleon source. The fireball radius $r_0$ and suppression radius $r_*$ are indicated by a  solid-line and a dashed-line respectively \cite{Kittiratpattana:2020daw}.}
    \label{fig:fig8}
\end{figure}

The result, shown in the figure \ref{fig:fig8}, supports the idea of (anti)nucleon suppression at higher energy. The increase of the fireball volume from UrQMD also agrees with the increase of the charged volume from thermodynamics. Furthermore, the suppression radius $r_*$ of the antinucleon source decreases at higher energy as expected. Compared to the coalescence model and charged volume constraint methods, the UrQMD result does not give any maximum below $\sqrt{s_{NN}}=200$~GeV. Also, the (anti)nucleon fireballs predicted by UrQMD are larger than resulted from the other methods, i.e., $7 \leq r_0 \leq 12$~fm. To compare results from all three methods, we look at the relative difference of the geometries.   

\begin{figure}
    \centering
    \includegraphics[width=0.9\columnwidth]{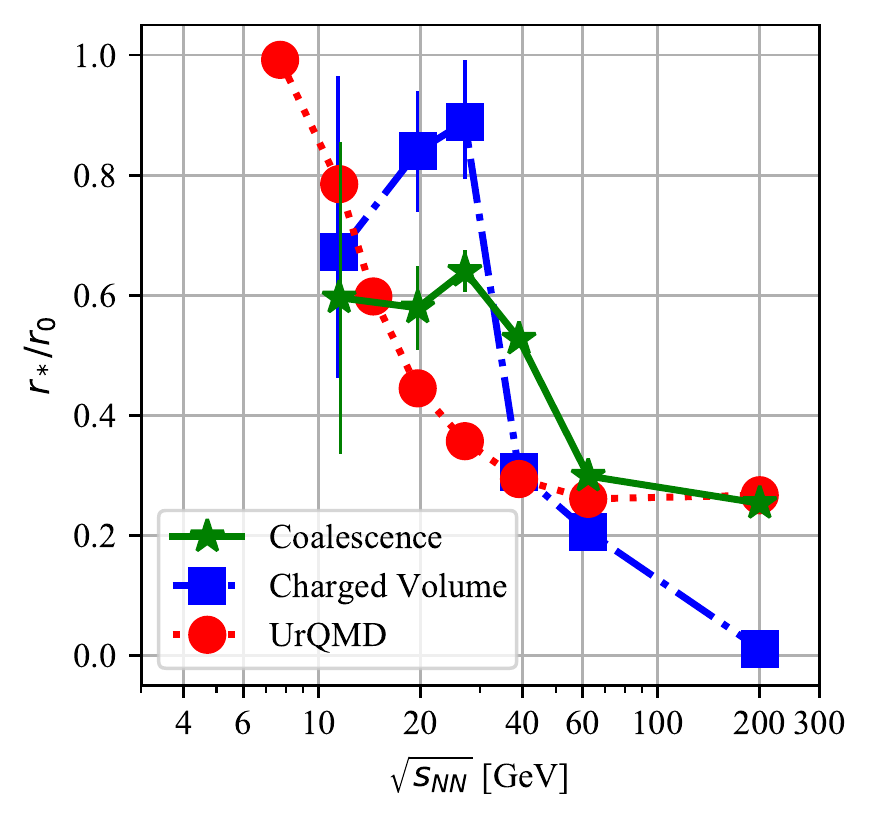}
    \caption{The $r_*/r_0$ ratio of antinucleon source from coalescence model (green star symbol), charged volume constrain (blue square symbol), and UrQMD simulation (red circle symbol). }
    \label{fig9}
\end{figure}

The $r_*/r_0$ ratio shows a common trend shared by the three methods (see figure~\ref{fig9}). At the lowest energy, we expect the ratio to be 1, since there is no pion dominance. Antinucleons should be suppressed the most. The UrQMD model (red circle symbol) gives us a satisfying result where $r_*/r_0$ becomes 1 at low energy. However, as we need more experimental data at this low energy, we cannot predict this expectation to the other method. At higher energies, the suppression region shrinks as pions get more abundant. The ratio at this energy range should drop rapidly to zero since there is no suppression caused by any nucleon-antinucleon annihilation. In this case, the charged volume RMS-radius constraint method (blue square symbol) fits the expectation very well. The spatial coalescence model (green star symbol) presents a behaviour of maximum suppression $r_*$ at $\sqrt{s_{NN}}=27$~GeV similar to the RMS-radius method. 


\section{Conclusion}
We applied the spatial coalescence model to study and extract the energy dependence of the (anti)nucleon source geometry, i.e., the fireball radius $r_0$ and suppression radius $r_*$ by fits to the measured coalescence parameter $B_2$ from NA49 and STAR experiments. We found that both, the nucleon and antinucleon source radii $r_0$, as well as antinucleon source suppression radius $r_*$, grow along with the  beam energy until $\sqrt{s_{NN}}=27$~GeV before decreasing at higher energies where the fireball becomes pion dominated. Hence nucleon-antinucleon becomes less important for the source geometry. However, this method did allow to understand how the nucleon source is affected by the pion domination because the suppression term of the nucleon source function could not be taken into account. Also, the decreasing nucleon fireball volume $r_0^3$ does not agree with the fireball volume expected from the number of charged particles.

Thus, we supplemented our model to include the volume effect for nucleons based on the charged particle yield via the RMS-radius of the source. We evaluated the antinucleon and nucleon source geometry $(r_0, r_*)$ pair by solving two equations simultaneously, i.e., the formation rate as a function of coalescence parameter $\mathcal{A}(r_0, r_*) = \frac{m_N}{2}B_2$, and the RMS-radius as a function of the number of charged particles $\langle r^2\rangle^{1/2}_{ch} = 0.925 \cdot N^{1/3}_{ch}$. In this method, both nucleon and antinucleon sources are independent of each other. This allowed us to observe how the nucleon source geometry is affected by the volume via the charged particle yield. The result showed that, at higher energies, the suppression region $r_*$ inside the nucleon source also decreases until it vanishes to zero, which means that a nucleon and anti-nucleons source become similar and a gain of a pure Gaussian form. 

Finally, we employed UrQMD to further confirm the freeze-out behaviour of nucleons and antinucleons in the transverse plane. The results show that both nucleon and antinucleon fireball radii increase monotonously, i.e., there is no peak at any energy. At lower energy, antinucleons suffer from the nucleon-antinucleon annihilation reflecting the deep valley near the center of the source. On the other hand, at higher energies, both nucleons and antinucleons behave similarly. In the case of antinucleons, the central suppression becomes smaller and similar to the nucleons, because now the dynamics is dominated by pions which suppress the freeze-out of deuterons and antideuterons equally. Quantitatively, this seems to be in contradiction to the result of charged constraint method where nucleon suppression vanishes  (i.e. $r_* \simeq 0$~fm for high energies). However qualitatively the results are comparable because the effective suppression as measured by the valley to peak ratio of the emission function in figure~\ref{fig:fig7} drops from $0.03/0.14\approx 0.2$ to $0.05/0.08=0.6$ indicating that the source becomes flatter in $r_T$.

Ultimately, all the methods support the interpretation of the antinucleon and nucleons source geometry being driven by $N\bar N$-absorption effects at low energies and by pions at high energies.  We also found a peculiar critical behaviour of the suppression radius $r_*$ around $\sqrt{s_{NN}}\approx 20$~GeV, which seems to hint at a longer lifetime of the system leading to stronger suppression of the anti-nucleons source. 


\section{Acknowledgement}
We thank for the support by the Development and Promotion of Science and Technology Talents Project (DPST) - Royal Thai Government Scholarship, Suranaree University of Technology (SUT), the Deutscher Akademischer Austausch Dienst (DAAD), the Stiftung Polytechnische Gesellschaft Frankfurt am Main, the Helmholtz International Center for FAIR (HIC for FAIR) within the LOEWE program launched by the State of Hesse, and the COST Action CA15213 (THOR). We wish to thank the organizers of the 9th International Conference on New Frontiers in Physics (2020) for publishing this abstract in their conference proceedings and the participants for fruitful discussions. Finally, we would also like to express our gratitude to our colleagues at the Frankfurt Institut f\"ur Advanced Studies (FIAS) and at SUT as well as Prof S. Mr\'owczy\'nski for valuable discussions and guidance.
\\


\section*{References}
\bibliographystyle{iopart-num}
\bibliography{mybib.bib}

\end{document}